\documentclass[conference]{IEEEtran}
\IEEEoverridecommandlockouts
\usepackage{cite}
\usepackage{amsmath,amssymb,amsfonts}
\usepackage{algorithmic}
\usepackage{graphicx}
\usepackage{textcomp}
\usepackage{siunitx}
\DeclareSIUnit\byte{B}
\DeclareSIUnit\tebibyte{TiB}
\DeclareSIUnit\gibibyte{GiB}
\usepackage{xcolor}
\usepackage{hyperref}
\usepackage[utf8]{inputenc}

\def\BibTeX{{\rm B\kern-.05em{\sc i\kern-.025em b}\kern-.08em
    T\kern-.1667em\lower.7ex\hbox{E}\kern-.125emX}}

\begin{document}

\title{Large-Scale Quantum Circuit Simulation on an Exascale System for QPU Benchmarking\\
}

\author{\IEEEauthorblockN{1\textsuperscript{st} J. A. Montanez-Barrera}
\IEEEauthorblockA{\textit{Jülich Supercomputing Centre} \\
\textit{Forschungszentrum Jülich}\\
Jülich, Germany \\
j.montanez-barrera@fz-juelich.de}
\and
\IEEEauthorblockN{2\textsuperscript{nd} Kristel Michielsen}
\IEEEauthorblockA{\textit{Jülich Supercomputing Centre} \\
\textit{Forschungszentrum Jülich}\\
Jülich, Germany}
}

\maketitle

\begin{abstract}
Recent advances in quantum computing have enabled the development of quantum processors with hundreds of qubits. However, noise continues to limit the amount of useful information that can be extracted from these systems, making it essential to identify the regime in which experimental outputs remain reliable. In this work, we benchmark Quantinuum Helios-1, a 98-qubit trapped-ion quantum processing unit, using the linear ramp quantum approximate optimization algorithm (LR-QAOA). To this end, we perform large-scale noiseless simulations on JUPITER, Europe’s first exascale supercomputer, for circuits of up to 48 qubits and 3,384 two-qubit gates. These simulations, executed on 4,096 nodes equipped with 16,384 GH200 superchips and high-bandwidth CPU–GPU interconnects, provide a reference for validating experimental results at the edge of classical tractability. We find that, up to 48 qubits, Helios-1 remains in a noise-tolerant region, i.e., its samples cannot be clearly distinguished from those coming from a noiseless simulation. We then extend the analysis to larger system sizes using experimental data only, and apply a mean-of-means resampling procedure with a $3\sigma$ threshold to determine whether the QPU output is statistically distinguishable from random sampling. This analysis identifies a regime of coherent performance up to 93 qubits (12,834 two-qubit gates), beyond which, at 95 qubits, the outputs become statistically indistinguishable from random sampling. These results demonstrate how exascale classical simulation can be used to validate quantum processors, and provide a quantitative boundary between noise-tolerant and random regimes in quantum processors.
\end{abstract}


\section{Introduction}
Recent advances in quantum computing have enabled the realization of processors with tens to hundreds of qubits \cite{Kim2023UtilityPreFault, ransford2025helios98qubittrappedionquantum, willow}. However, noise and hardware imperfections continue to limit the reliability of quantum computations, making it essential to determine when and to what extent experimental outputs are reliable. A central challenge is to identify the boundary between regimes where quantum processors exhibit coherent, algorithmically meaningful behavior and those where noise renders outputs effectively random.

Existing benchmarking approaches span a broad range of methodologies~\cite{lall2025reviewcollectionmetricsbenchmarks, Hashim_2025}, from component-level metrics such as randomized benchmarking~\cite{Emerson_2005} and error per layered gate (EPLG)~\cite{mckay2023benchmarkingquantumprocessorperformance}, to system-level tests like quantum volume (QV)~\cite{PhysRevA.100.032328} and cross-entropy benchmarking (XEB)~\cite{Boixo2018}. While these protocols are valuable for characterizing gate fidelities and small-scale circuit execution, many rely on classical simulation of ideal output distributions or evaluate sections of a quantum processing unit (QPU) in a disjoint way which limits their general evaluation. Application-level benchmarks based on the linear ramp quantum approximate optimization algorithm (LR-QAOA) have recently been introduced as a scalable complement to these methods~\cite{montanezbarrera2025}. LR-QAOA enables evaluation of QPUs across both circuit width and depth using a unified, platform-agnostic framework. The key performance metric is the approximation ratio $r$, which increases with depth and saturates at 1 in the absence of noise, degrading as coherence is lost. It assesses performance through statistical comparison with random sampling, offering algorithmic-level insight at scales where simulation-dependent methods can no longer be applied. This method has already been adopted in practical benchmarking frameworks by both industry and non-profit initiatives, including work by IonQ~\cite{aboumrad2026measuringmattersscalableframework} and the Unitary Foundation~\cite{cosentino2026metriqcollaborativeplatformbenchmarking}.

In this work, we use LR-QAOA to benchmark Quantinuum’s Helios, a 98-qubit trapped-ion QPU \cite{ransford2025helios98qubittrappedionquantum}. Large-scale noiseless simulations on JUPITER \cite{deraedt2025universalquantumsimulation50}, Europe’s first exascale supercomputer, allow us to validate the QPU’s performance comparison with an ideal case for up to 48 qubits. Experimental benchmarking is extended up to 98 qubits, beyond the limits of classical verification. Our results show that in this regime, still some coherent performance is obtained up to 93 qubits (12,834 two-qubit gates), beyond which outputs become statistically indistinguishable from random, establishing a quantitative boundary between noise-tolerant and noise-dominated operation. This work demonstrates how exascale classical simulation can be used to validate quantum processors and provides a scalable, interpretable framework for algorithmic-level benchmarking as quantum devices approach the limits of classical tractability.

The main contributions of this work are:
\begin{enumerate}
    \item First quantum circuit simulation on an exascale supercomputer for QPU benchmarking: noiseless simulations on JUPITER using up to 4,096 nodes (16,384 GH200 superchips) for circuits up to 48 qubits (3,384 two-qubit gates). The 48-qubit simulation represents the largest reported QAOA simulation at FP32 precision to date.
    \item Coherence boundary identification: using noiseless JUPITER simulations, Helios-1 is certified to operate in a noise-tolerant regime up to 48 qubits; extending experimentally to 98 qubits, some coherent performance is maintained up to 93 qubits (12,834 two-qubit gates), beyond which outputs become indistinguishable from random sampling.
    \item Low-shot statistical methodology for QPU performance evaluation: mean-of-means resampling with a $3\sigma$ threshold classifies QPU outputs into noise-tolerant, transition, or random regimes with as few as 10 shots.
    \item Cross-platform GPU benchmarking for simulating quantum circuits: H100 achieves ~1.9× speedup over A100 on a 30-qubit simulation and JUPITER matches JUWELS Booster execution time of a 40-qubit simulation using half the GPUs.
\end{enumerate}

The remainder of the paper is organized as follows.
Section~\ref{Sec:Methods} introduces the benchmarking framework, beginning with a survey of existing protocols and the LR-QAOA benchmark used in this work, followed by a noise model that motivates the statistical test we apply. We then describe the sampling methodology, the classical simulation infrastructure on JUPITER, the quantum processor under test, and the experimental setup. In Sec.~\ref{Sec:Results}, we present simulation performance results and QPU benchmarking outcomes. Finally, Sec.~\ref{Sec:Conclusions} summarizes the main conclusions.

\section{Methods}\label{Sec:Methods}

\subsection{Benchmarking Quantum Computers} 

The landscape of quantum benchmarking protocols has grown substantially in recent years, spanning component-level, system-level, and application-level approaches~\cite{lall2025reviewcollectionmetricsbenchmarks, Hashim_2025}. At the component level, randomized benchmarking (RB)~\cite{Emerson_2005, Helsen_2022} estimates average gate fidelities through sequences of random Clifford operations; however, standard RB targets individual or small groups of qubits. Error per layered gate (EPLG)~\cite{mckay2023benchmarkingquantumprocessorperformance} extends this by measuring error rates across disjoint layers of two-qubit gates over a full chip.

At the system level, quantum volume (QV)~\cite{PhysRevA.100.032328} provides a single-number metric by testing the largest square circuit a device can execute reliably via the heavy-output generation problem. While widely adopted, QV requires classical simulation of the output distribution, limiting it to roughly 50 qubits~\cite{Baldwin_2022}, and independent studies have shown that manufacturer-reported QV values are often difficult to reproduce~\cite{Pelofske_2022}. Cross-entropy benchmarking (XEB)~\cite{Boixo2018}, used in quantum supremacy demonstrations is also presented as a benchmarking alternative. The recently proposed Clifford volume~\cite{portik2025cliffordvolumefreefermion} addresses the scalability gap by using efficiently verifiable stabilizer circuits.

Application-level benchmarks evaluate quantum processors through structured problem instances rather than synthetic circuits. The Q-score~\cite{Martiel_2021} measures the largest MaxCut instance for which a QPU can find the optimal solution of the problem, and the concept of algorithmic qubits (AQ)~\cite{Chen2023} captures effective qubit count for a given task. While these benchmarks provide useful insights, they have limitations. Q-score distinguishes QPU performance from random sampling, but depends on parameter tuning, reflecting both hardware execution and algorithm optimization, which complicates cross-platform comparisons. AQ provides a volumetric view across qubits and depth, but is also sensitive to compilation and error mitigation. As a result, both approaches do not fully isolate intrinsic QPU performance.

In this work, we use the LR-QAOA-based benchmark~\cite{montanezbarrera2025}, which evaluates QPU performance as a function of both circuit width and depth on combinatorial optimization instances. Because LR-QAOA uses a deterministic, non-variational protocol, it is straightforward to implement and reproduce across platforms. Its validation requires comparing the QPU's approximation ratio $r$ against random baselines. The fully connected graph instances used here produce dense circuits in which every qubit interacts with every other, stressing the QPU under conditions where correlated errors and all-to-all communication are unavoidable. This complements existing system-level and application-level benchmarks by providing algorithmic-level insight while isolating intrinsic QPU performance from optimization and mitigation effects, even at problem sizes beyond classical simulability. The following subsection details the LR-QAOA protocol and its key quantities.

\subsection{LR-QAOA Benchmarking Protocol}
LR-QAOA is a non-variational version of QAOA~\cite{Farhi2014} that uses a linear annealing schedule for its parameters, thereby removing the need for classical optimization. It can be interpreted as a first-order Trotterized approximation of an adiabatic quantum evolution~\cite{farhi2000quantum}. The algorithm consists of alternating applications of a problem Hamiltonian, whose ground state encodes the solution to a combinatorial optimization problem (COP), and a mixer Hamiltonian, whose ground state is used as the initial state of the evolution. As the system evolves through $p$ layers, the average energy decreases, amplifying high-quality solutions. In this work, we focus on the fully connected Weighted Max-Cut (WMC) problem due to its formulation in terms of only two-qubit interactions, making it well suited for implementation on quantum hardware.

The WMC cost Hamiltonian is defined as
\begin{equation}
H_C = \sum_{\{i, j\} \in E(\mathcal{G})} w_{ij} \sigma_z^i \sigma_z^j, 
\end{equation}
where $\mathcal{G}$ is a weighted graph, $w_{ij}$ are edge weights, and $\sigma_z^i$ denotes the Pauli-$z$ operator acting on qubit $i$, which represents a vertex of the graph. The corresponding unitary operator applied in each layer is
\begin{equation}\label{UC}
    U_C(H_C, \gamma_k)=e^{-j \gamma_k H_C},
\end{equation}
with $\gamma_k$ determined by the linear ramp schedule and $j=\sqrt{-1}$. This is followed by the application of the mixer unitary
\begin{equation}\label{UB}
    U(H_B, \beta_k)=e^{j \beta_k H_B},
\end{equation}
where $\beta_k$ is also set by the linear schedule and $H_B = \sum_{i=0}^{N_q-1} \sigma_i^x$. The full circuit consists of $p$ repetitions of these unitaries, starting from the initial state $|+\rangle^{\otimes N_q}$. The number of gates needed to implement this protocol on a fully connected QPU are $N_{2q} = p N_q(N_q-1)/2$ two-qubit gates and $N_{1q} = (p+1) N_q$ single qubit gates for $N_q$ qubit.

 Repeated preparation and measurement of the final LR-QAOA state produces a distribution over candidate solutions, whose quality improves with circuit depth in the noiseless case. The protocol is characterized by three parameters $\Delta_\beta$, $\Delta_\gamma$, and the number of layers $p$. The parameters $\beta_i$ and $\gamma_i$ follow the linear schedules
\begin{equation}
\beta_i = \left(1-\frac{i}{p}\right)\Delta_\beta\ \ \mathrm{and} \ \
\gamma_i = \frac{i+1}{p}\Delta_\gamma,
\end{equation}
for $i=0, \dots, p-1$. We use the approximation ratio as the performance metric, it is described by 
\begin{equation}
r = \frac{\sum_{i=1}^{n_s} C(x_i)/n_s}{C(x^*)},
\label{eq:r}
\end{equation}
where the cost function is defined as
\begin{equation}
C(x) = \sum_{k,l>k}^{N_q} w_{kl}(x_k + x_l - 2x_kx_l).
\end{equation}
Here, $n_s$ is the number of samples, $x_i$ denotes the $i^{\mathrm{th}}$ bitstring obtained from LR-QAOA, $x^*$ is the optimal bitstring, and $C(x^*)$ is the maximum cut value. By comparing a QPU's approximation ratios against noiseless and random baselines, one can identify the transition between coherent and noise-dominated regimes. In practice, noise in QPUs degrades LR-QAOA performance as circuit depth and width increase; we characterize this effect next.

\subsection{Noise in LR-QAOA}
At the gate level, the dominant source of noise in QPUs comes from two-qubit entangling gates~\cite{Pascuzzi2022}. To study this effect within the LR-QAOA protocol, we model noise by applying a depolarizing channel to all two-qubit gates. For simplicity, we assume a uniform depolarizing error rate $\varepsilon$ across all two-qubit gates.

\begin{figure}[htbp]
\centering
\includegraphics[width=1\linewidth]{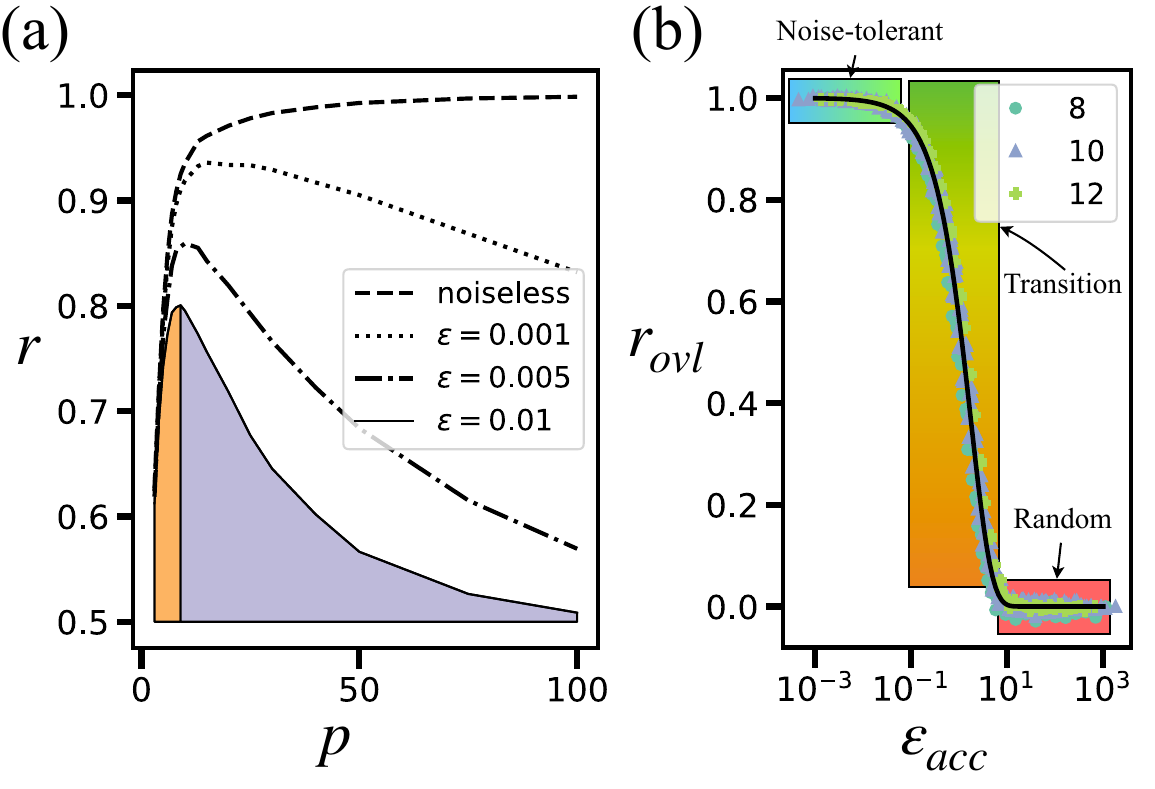}
\caption{Noise in LR-QAOA. (a) Simulation of LR-QAOA under different levels of depolarizing noise as a function of circuit depth. (b) Overlap between the noisy and noiseless approximation ratios for 8-, 10-, and 12-qubit simulations. The solid lines correspond to the prediction from Eq.~\ref{Eq:noise}. Three algorithm performance regimes are highlighted with boxes.}
\label{fig:noise}
\end{figure}

Figure~\ref{fig:noise}(a) shows a simulation of LR-QAOA behavior of $r$ as a function of the number of layers $p$ under different levels of depolarizing noise $\varepsilon$. In the noisy regime, the evolution is separated into two regimes, the first, highlighted in orange, is dominated by LR-QAOA performance, where the algorithmic signal increases faster than noise accumulates; the second, shown in purple, is noise-dominated, with $r$ eventually approaching the random-sampling limit. At intermediate $\varepsilon$, performance increases consistently towards the  noiseless simulation. For reference, the noiseless case (dashed line) grows monotonically towards the maximum $r=1$. 

In~\cite{Monta_ez_Barrera_2025}, it is shown that the performance of a noisy QPU can be quantified using the normalized overlap between the ideal evolution and the QPU evolution:  

\begin{equation}
    r_{ovl} = \frac{r_{\mathrm{QPU}} - r_{\mathrm{random}}}{r_{\mathrm{ideal}} - r_{\mathrm{random}}},
\end{equation}
where $r_{ovl}$ measures the overlap between the ideal approximation ratio $r_{\mathrm{ideal}}$ and the experimentally obtained value $r_{\mathrm{QPU}}$, normalized by the random-sampler approximation ratio $r_{\mathrm{random}}$. This overlap can be related to the accumulated error, defined as $\varepsilon_{\mathrm{acc}} = N_{2q} \varepsilon$, via  

\begin{equation}\label{Eq:noise}
    r_{ovl} = 2^{-k_0 \varepsilon_{\mathrm{acc}}},
\end{equation}
where $k_0$ is a problem-dependent fitting parameter. Figure~\ref{fig:noise}(b) shows that this model accurately describes the effect of simulated depolarizing noise for different problem sizes and LR-QAOA depth. The solid lines correspond to the fits of Eq.~\ref{Eq:noise}, while the markers indicate results for various system sizes. We identify three distinct regions: the {\it noise-tolerant region}, where the accumulation of noise is still negligible and performance is largely preserved meaning a $r_{ovl}\approx 1$; the {\it transition region}, where noise begins to reduce performance but LR-QAOA signals are still distinguishable; and the {\it random region}, where outputs become indistinguishable from a random sampler, corresponding to a fully mixed state in the QPU. Although derived from a simplified depolarizing-noise model, this behavior has been shown to consistently reproduce QPU performance, as demonstrated in \cite{Monta_ez_Barrera_2025}. To determine whether experimental outputs fall within the noise-tolerant, transition, or random regime, we require both a noiseless classical reference and a statistical test. We describe the statistical methodology next, followed by the classical simulation infrastructure used to produce noiseless references.

\subsection{Sampling Methodology}
The three performance regimes identified above, noise-tolerant, transition, and random, must be distinguished experimentally from a finite number of QPU samples. We now describe the statistical procedure used for this classification.

Given a QPU output, we compare its approximation ratio against two reference distributions: one constructed from a random sampler (uniform bitstrings evaluated on the same graph) and, where available, one from noiseless classical simulation. The QPU result is classified as \textit{noise-tolerant} if it falls within the 99.73\% confidence interval of the noiseless distribution, in the \textit{transition} regime if it lies between the noiseless and random intervals, and in the \textit{random} regime if it is within or below the random interval.

To quantify the statistical separation between QPU output and these baselines, we construct kernel density estimates (KDEs) of the mean approximation ratio $\bar{r}$ obtained by repeatedly subsampling from the reference distributions.
Given a sample size $n_s$, we draw $n_s$ bitstrings uniformly at random (without replacement) from the full output distribution, compute the mean approximation ratio $\bar{r}_i$ for each subset $i$, and repeat 100 times.
This yields an empirical distribution of subsample means $\{\bar{r}_i\}_{i=1}^{100}$, from which we compute the grand mean $\langle \bar{r} \rangle = \frac{1}{100}\sum_i \bar{r}_i$ and the standard deviation $\sigma_{\bar{r}}$.
The random threshold is then defined as $\langle \bar{r} \rangle + 3\sigma_{\bar{r}}$, and the width of this mean-of-means distribution reflects the sampling uncertainty at a given number of shots.

A QPU result is classified as statistically meaningful if the observed approximation ratio $\bar{r}_{\mathrm{QPU}}$ exceeds the random threshold $\langle \bar{r} \rangle + 3\sigma_{\bar{r}}$, i.e., with a 99.73\% confidence.

\begin{figure}[htbp]
\centering
\includegraphics[width=1\linewidth]{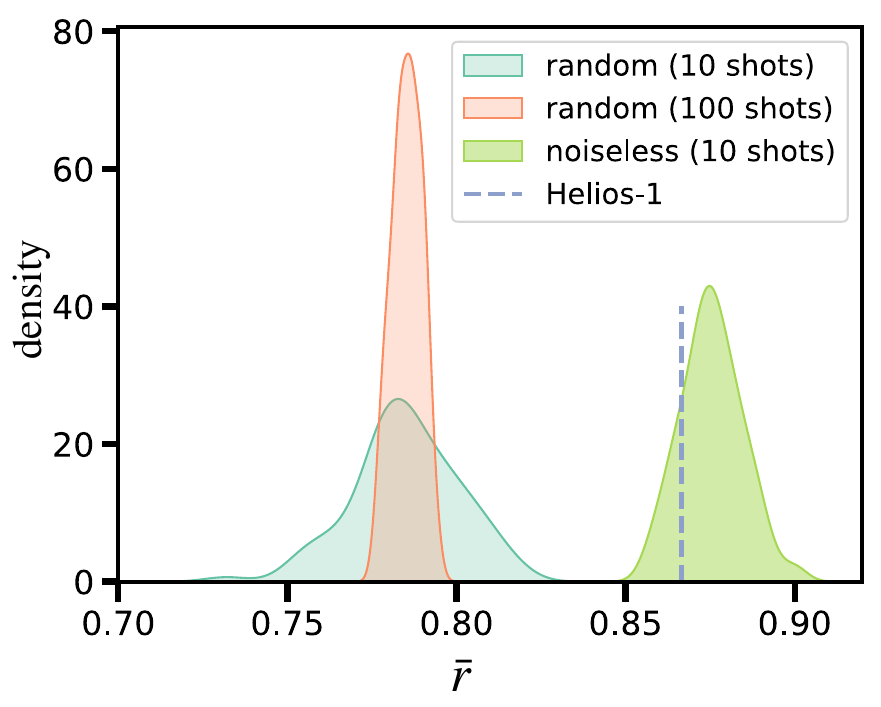}
\caption{Sampling distribution for the 40-qubit LR-QAOA benchmark at $p=3$. KEDs of the mean approximation ratio $\bar{r}$ are shown for the random sampler at 10 and 100 shots, and for the noiseless simulation at 10 shots. The dashed line indicates the Helios-1 QPU result.}
\label{Fig:sampling}
\end{figure}

Figure~\ref{Fig:sampling} illustrates this procedure for the 40-qubit fully connected problem on Helios-1 at $p=3$.
At 100 shots, the random sampling distribution is narrowly concentrated around $\bar{r}_{\mathrm{rand}} \approx 0.79$ with a $3\sigma$ upper bound of $\approx 0.80$, whereas at 10 shots the distribution broadens considerably, the standard deviation increases by a factor of $\approx 3.5$. This highlights a trade-off inherent to the small-sample regime: while 100 samples tightly constrain the random baseline, in scenarios where the cost of sampling is significant, it is desirable to observe the separation between algorithmic signal and noise even with few samples.

The noiseless (JUQCS simulation on JUPITER) distribution at 10 shots is centered at $\bar{r}_{\mathrm{ideal}} \approx 0.88$, well separated from the random baseline. The QPU result from Helios-1 ($\bar{r} \approx 0.87$) falls squarely within the noiseless region and far above the random $3\sigma$ threshold, confirming a statistically meaningful algorithmic signal. Computing the noiseless reference $\bar{r}_{\mathrm{ideal}}$ required for this test demands large-scale quantum circuit simulation, which we describe next.

\subsection{Classical Simulation Setup}

State-vector simulation is the standard approach to represent the full quantum state as a complex vector of dimension $2^N$, allowing exact simulations but at high resource cost. Examples of this approach are IBM Qiskit AerSimulator~\cite{Qiskit}, Google Cirq~\cite{Cirq}, Eviden's Qaptiva~\cite{Qaptiva}, ScaleQsim \cite{Qsim}, and JUQCS \cite{RAED19a,Willsch2020,deraedt2025universalquantumsimulation50}. From them, JUQCS is the only one shown to simulate quantum circuits larger than 42 qubits due to its memory and processing capabilities. JUQCS runs across diverse hardware platforms, from desktop PCs to high-end supercomputers with distributed or shared memory, requiring only a Fortran 2003-compatible compiler and MPI support on CPUs, and CUDA-Fortran plus CUDA-aware MPI for NVIDIA GPUs.

JUQCS simulation of a universal $N$-qubit quantum computer requires storing the full state vector. It has three precision options: BE (byte), single precision (FP32), and double precision (FP64). Throughout this work, we use FP32 as we did not notice a difference between FP32 and FP64 in the outcome of our simulations, and small-scale experiments showed that BE does not perform well on this task. FP32 demands $2^{N_q+3}$ bytes of memory~\cite{RAED07x}. Each quantum gate updates the state vector by multiplying disjoint pairs (for single-qubit gates) or quadruples (for two-qubit gates) of elements by the corresponding $2\times2$ or $4\times4$ matrix. These operations are inherently parallel, but as the state vector grows beyond the memory of a single processing unit, it must be distributed across multiple devices, in JUPITER, NVIDIA Grace Hopper GH200 superchips~\cite{deraedt2025universalquantumsimulation50}. If $N_q'$ qubits fit on a single superchip, $n = 2^{N_q-N_q'}$ superchips are needed, each storing $L=2^{N_q'}$ elements. Gates acting only on qubits with indices $0\le j<N_q'$ can be applied independently on each superchip without communication, while gates targeting qubits $j\ge N'$ require redistribution of half of the state-vector elements between pairs of superchips. This communication pattern, determined by the target qubits, is a key factor in the efficiency of large-scale state-vector simulations.

Simulating a 48-qubit quantum computer in FP32 precision on a GH200-based system requires storing $2^{48}$ complex amplitudes, corresponding to approximately \SI{2,048}{\tebibyte} of memory. JUPITER provides up to 6,000 nodes (\num{24,000} superchips), but JUQCS-50 can utilize at most 4,096 nodes (\num{16,384} superchips) due to powers-of-two constraints. Each GH200 superchip offers \SI{96}{\gibibyte} of device memory and \SI{120}{\gibibyte} of host memory. Using only the GH200 device memory, a single chip can store up to 33 qubits (\SI{64}{\gibibyte}), while utilizing the combined device and host memory allows simulation of 34 qubits (\SI{128}{\gibibyte}). For instance, the 46-qubit simulation using device memory alone requires $n = 2^{46-33} = 8,192$ chips, and in the 48-qubit simulation using the extended memory, $n=2^{48-34} = 16,384$ chips are needed.
However, using the combined device and host memory incurs a significant computational overhead: for example, the 44-qubit at $p=3$ LR-QAOA simulation using only GPU memory takes approximately 418\,s, whereas employing the extended memory increases the runtime to 1,818\,s. These noiseless simulations provide the reference baselines against which we compare outputs from the quantum processor described next.

\begin{figure*}[t!]
\centering
\includegraphics[width=1\linewidth]{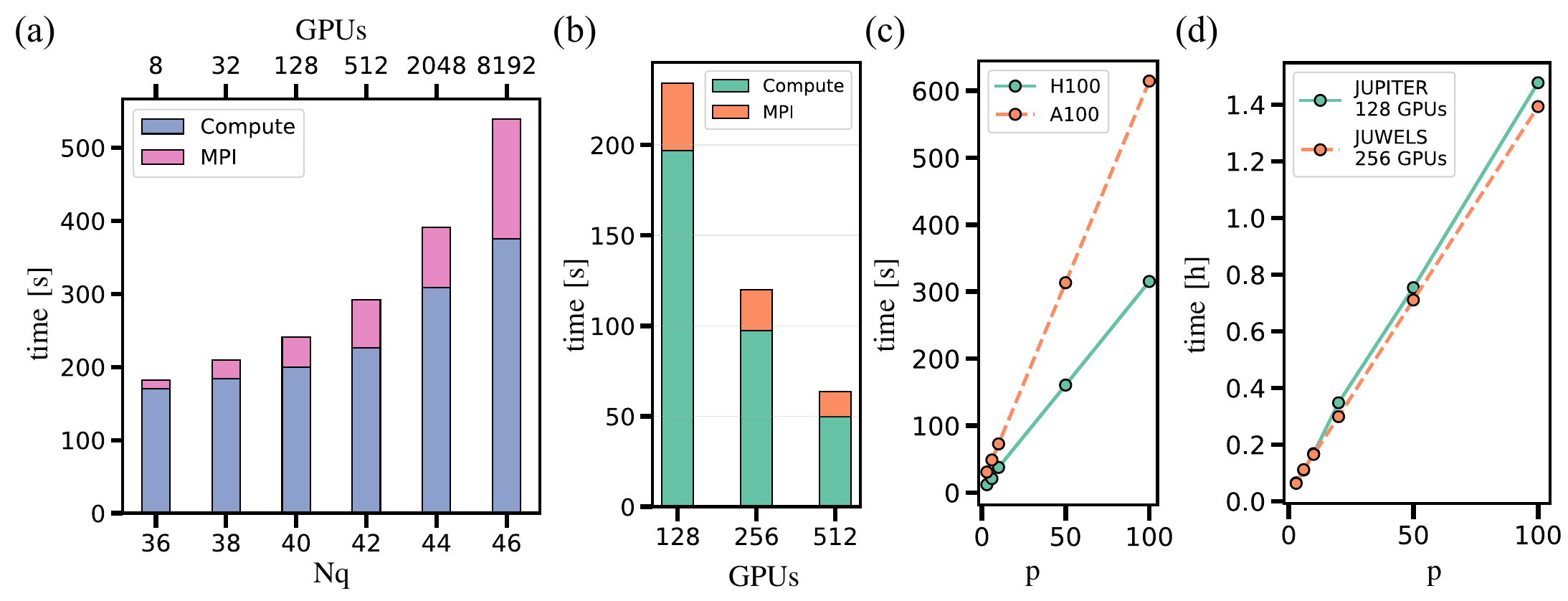}
\caption{Time scaling of LR-QAOA circuits on JUPITER. (a) Runtime as a function of problem size. (b) Strong scaling for the 40-qubit problem. (c) Hardware performance comparison between H100 (JUPITER) and A100 (JUWELS Booster) GPUs. (d) Performance comparison between JUPITER and JUWELS Booster for a 40-qubit problem as a function of LR-QAOA depth.}
\label{fig:scaling}
\end{figure*}

\subsection{Quantum Processor}
The experiments in this work are performed on the new generation of Quantinuum trapped-ion quantum processor, Helios-1, a 98-qubit system based on a quantum charge-coupled device (QCCD) architecture \cite{ransford2025helios98qubittrappedionquantum}. The device uses $^{137}$Ba$^{+}$ hyperfine qubits and enables effectively all-to-all connectivity through ion shuttling between multiple operation zones connected by a junction. Quantum circuits are executed using a combination of single-qubit rotations and two-qubit entangling gates, with parallelized operations to improve execution speed. The control stack includes real-time compilation and dynamic circuit execution, allowing adaptive program flow during runtime. Helios-1 reports a two-qubit gate infidelity of approximately $7.9 \times 10^{-4}$~\cite{ransford2025helios98qubittrappedionquantum}, which determines the noise budget across the circuits studied here.

On Helios-1, executed jobs are charged in Hardware Quantum Credits (HQCs). Because the processor supports conditional branching, the total HQC cost is dynamic and depends on the specific branches executed at runtime, plus a fixed overhead of 5~HQCs per job submission~\cite{quantinuum_workflow}. The cost is approximated by
\begin{equation}
\mathrm{HQC} = 5 + \frac{N_{1q} + 10\,N_{2q} + 5\,N_m}{5000}\,n_s,
\end{equation}
where $N_m$ denote the number of measurements. Because $N_{2q}$ grows quadratically with problem size for the benchmarking, the credit cost also scales quadratically. For example, executing the $N_q=40$ benchmark at $p=3$ with $n_s=10$ shots requires $\mathrm{HQC}\approx 68$, while $N_q=98$ requires $\mathrm{HQC}\approx 387$. The full suite of twelve experiments used in this work, spanning 40 to 98 qubits, costs roughly $\mathrm{HQC}\approx 2{,}460$ at $n_s=10$. Repeating the same suite at $n_s=100$ would require approximately $\mathrm{HQC}\approx 24{,}000$, a budget that might exceed the allocations available to many research groups and that would grow further for deeper circuits or denser sampling strategies commonly used in other benchmarks. This cost structure shows a practical advantage of LR-QAOA; as shown in Sec.~\ref{Sec:Methods} and Fig.~\ref{Fig:sampling}, the protocol yields statistically meaningful separation between the QPU signal and the random baseline with as few as 10~shots.

\subsection{Experimental Setup}
We evaluate Helios-1 using the LR-QAOA benchmark at depth $p=3$ with $\Delta_{\beta}$ and $\Delta_{\gamma}=0.2$ on fully connected weighted MaxCut instances. For each problem size $N_q$, a single random graph instance is generated with edge weights drawn uniformly from $[0,1]$; the optimal cut value is computed with CPLEX. The number of qubits ranges from 40 to 98, covering both the classically verifiable regime ($N_q \le 48$, where noiseless JUPITER simulations are available) and the randomly verifiable regime ($N_q > 48$). On Helios-1, each circuit was executed with shots ranging from 9 to 49 across the different problem sizes. This choice is dictated by the budget constraints of device utilization, as the cost is primarily determined by the number of two-qubit gates in the experiment. The statistical test described in Sec.~\ref{Sec:Methods} is applied to every experiment to classify the QPU output into the noise-tolerant, transition, or random regime. For the 40-qubit case, we additionally compare Helios-1 results with those from the previous-generation H2-1 processor\cite{Moses_2023}.

With the benchmarking protocol and simulation infrastructure in place, we now present the results.

\section{Results}\label{Sec:Results}

\subsection{Quantum Simulation on JUQCS}

To evaluate the parallel performance of our implementation on the JUPITER supercomputer, we conducted strong scaling and problem-size analyses for the LR-QAOA algorithm with $p=3$ layers. 
In the problem-size scaling regime, Fig.~\ref{fig:scaling}(a), we increased both problem size from 36 to 46 qubits, which requires increasing the computational resources proportionally from 128 to 8192 GPUs (following $2^{n_q-33}$), revealing that execution time grows moderately from $\sim$182\,s to $\sim$540\,s. At larger scales (44-46 qubits with 2,048-8,192 GPUs), MPI communication overhead becomes increasingly significant, accounting for up to 30\% of total runtime.

Figure~\ref{fig:scaling}(b) shows the strong scaling at 40 qubits. We observed near-ideal speedup when increasing from 128 to 512 GPUs, achieving a $3.7\times$ reduction in total execution time (234\,s to 64\,s), with compute time consistently dominating MPI communication overhead across all configurations.

Additionally, in Fig.~\ref{fig:scaling}(c) we compared the single-GPU performance of NVIDIA H100 (JUPITER) and A100 (JUWELS Booster \cite{BOOSTER}) GPUs for a 30-qubit system across varying circuit depths ($p \in \{3, 6, 10, 50, 100\}$). The H100 consistently outperforms the A100, with the performance gap widening at deeper circuits: at $p=3$, both GPUs complete in under 10\,s, while at $p=100$, the H100 achieves $\sim$320\,s compared to $\sim$620\,s on the A100, representing a $\sim$1.9$\times$ speedup. This near-twofold improvement reflects the H100's enhanced compute, and is particularly relevant as circuit depth scales linearly with simulation time.

At 40 qubits, we further compared the total wall time per QAOA iteration on JUPITER (128 H100 GPUs) against JUWELS Booster (256 A100 GPUs), where both systems exhibit comparable performance, at $p=100$, JUPITER completes in $\sim$1.39\,h while JUWELS requires $\sim$1.38\,h, despite JUPITER using half the number of GPUs. This demonstrates that the superior per-GPU performance of the H100 effectively compensates for the reduced device count, requiring similar time with fewer computational resources.

\begin{figure*}[t]
\centering
\includegraphics[width=1\linewidth]{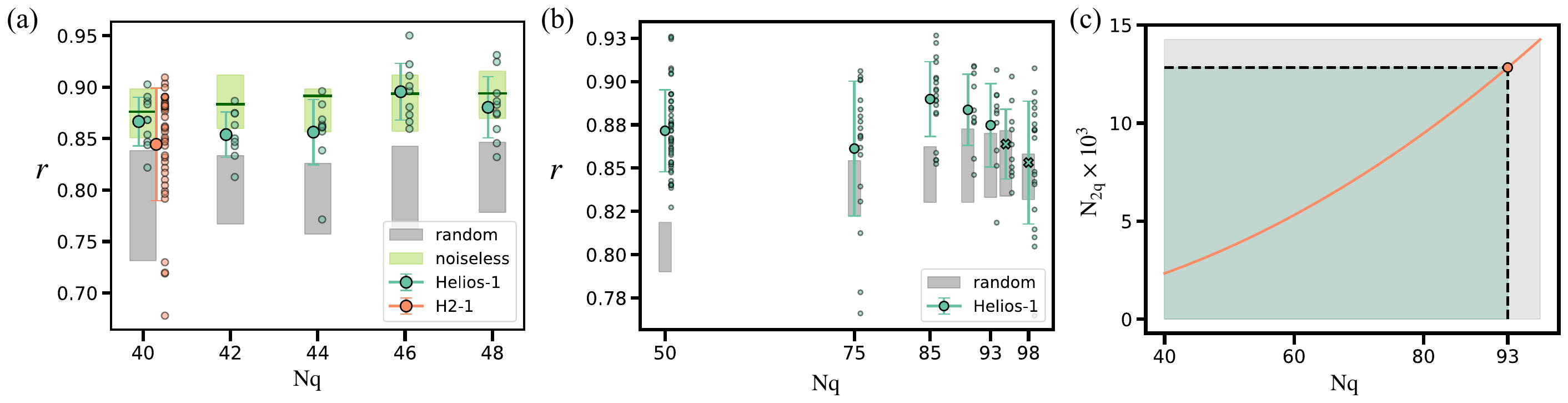}
\caption{LR-QAOA (p = 3) benchmarking of Helios-1 approximation ratio vs. number of qubits. (a) Comparison with noiseless classical simulations on JUPITER and with the previous-generation Quantinuum QPUs (H2-1). (b) Helios-1 experiments for problem sizes beyond the reach of exact JUQCS simulations on JUPITER. The solid dark green line indicates the true approximation ratio $r$.(c) Number of two-qubit gates as a function of problem size. The green shaded region and dashed line indicate the largest system size for which Helios-1 samples remain statistically distinguishable from random sampling. }
\label{fig:1}
\end{figure*}
\subsection{Benchmarking Helios-1} 
We benchmark Helios-1 quantum processor performance using LR-QAOA at depth p = 3 on fully connected weighted MaxCut instances across 40–98 qubits, covering both classically verifiable and intractable regimes. For 40-48 qubits, Fig.~\ref{fig:1}(a), results are validated against noiseless simulations using JUQCS on JUPITER to give a direct quantification of noise effects and comparison across hardware platforms (Helios-1 and H2-1). In this regime, the mean approximation ratio of Helios-1 samples lies within the 99.73\% confidence interval of the noiseless reference (JUPITER), and well above the random sampling threshold, i.e, in the {\it noise-tolerant regime}. This does not imply error-free circuit execution, but rather that the accumulated noise is insufficient to distinguish the QPU output from a noiseless simulation. Additionally, we include results from H2-1, the previous-generation Quantinuum processor, obtained with 50 samples at 40 qubits. Although the different sample sizes preclude a direct comparison, H2-1 benefits from lower sampling variance due to its larger shot count, Helios-1 achieves a comparable or higher approximation ratio with only 10 shots, suggesting an improvement in QPU performance across hardware generations.

The 48-qubit simulation represents the frontier of this classical verification effort. As described in Sec.~\ref{Sec:Methods}, this case requires using the combined device and host memory of every GH200 superchip, using \num{16,384} superchips across 4,096 JUPITER nodes. The resulting simulation took $\sim$ 2490\,s, nearly 4 times longer than the 46-qubit case due to the overhead of host-device data transfers. At this scale, the simulation also stresses the network infrastructure of JUPITER, with the nodes active simultaneously, the interconnect must sustain the redistribution of state-vector elements over an extended period.

For larger problem sizes (50–98 qubits), where classical verification is no longer feasible, performance is assessed relative to a random sampling baseline. We observe a gradual degradation with increasing system size, while the QPU maintains a statistically significant separation from the random baseline up to the 93-qubit experiment. Beyond this point, the 95- and 98-qubit outputs fall below the $3\sigma$ random threshold and are thus classified in the random regime. Coherent performance up to 93 qubits, corresponding to 12,834 two-qubit gates on a fully connected topology, shows that the QPU can still preserve a meaningful signal through deep, highly entangling circuits close to its maximum qubit capacity.

\section{Conclusions}\label{Sec:Conclusions}
In this work, we applied the LR-QAOA benchmarking protocol to evaluate the performance of the Helios-1 quantum processor. Large-scale simulations on JUPITER, Europe’s first exascale supercomputer, were performed for circuits up to 48 qubits, utilizing 4,096 nodes and 16,384 GH200 superchips. These simulations allowed us to certify that Helios-1 operates in a noise-tolerant regime up to 48 qubits, where noise has a minimal impact on LR-QAOA performance. We then extended the benchmarking experimentally on Helios-1 for LR-QAOA at $p=3$, scaling from 50 to 98 qubits. While ideal reference simulations become intractable at this scale, comparing QPU outputs to random distributions shows that the algorithm continues to produce meaningful results, with some coherent performance maintained up to 93 qubits, corresponding to 12,834 two-qubit gates. Beyond this, at 95 and 98 qubits, the QPU outputs fall into the random sampling regime, indicating the practical limits of current device coherence.

We also evaluated classical simulation performance, focusing on MPI communication and computation time for circuits ranging from 36 to 46 qubits. Computation time grows moderately from 170 to 375\,s despite a quadratic increase in operations, while MPI communication grows more sharply from 11 to 164\,s, corresponding to factors of 2.2 and 14.6, respectively. Strong scaling tests for the 40-qubit case demonstrate a time reduction of approximately 3.7$\times$ when using four times more GPUs than the minimum required. Furthermore, LR-QAOA enables comparisons across GPU generations: for a 30-qubit problem, NVIDIA H100 GPUs complete a $p=100$ simulation in 320\,s, nearly half the time of A100 GPUs, while 40-qubit simulations at $p=3$ achieve similar runtimes across JUWELS Booster and JUPITER, with JUPITER requiring only half the number of GPUs. These results highlight LR-QAOA as a versatile tool for benchmarking QPUs, assessing algorithmic depth under noise, and evaluating the efficiency of classical simulation platforms across hardware generations.

Overall, our work demonstrates that LR-QAOA provides a scalable, interpretable, and implementation-friendly framework for connecting algorithmic performance to device limitations. While the benchmarking relies on a limited number of samples for large circuits and uses moderate circuit depths ($p=3$), the protocol effectively identifies the transition between noise-tolerant and noise-dominated regimes. Compared to other quantum benchmarks, LR-QAOA directly links algorithmic performance to meaningful optimization metrics without relying on tuning or post-processing, making it particularly suitable for evaluating near-term QPUs. Future work will explore increasing the circuit depth to further stress-test coherence limits and extending the benchmarking framework to quantum error correction primitives.

\section*{Data Availability}
All problem instances and results analyzed in this study are available at: \url{https://jugit.fz-juelich.de/qip/lrqaoa-exascale-qpu-benchmarking}.

\section*{Acknowledgment}

The authors thank Hans De Raedt for the insightful discussions and suggestions made for the present work. J. A. Montanez-Barrera acknowledges support from the project EPIQ funded by MKW-NRW.
The authors gratefully acknowledge the Gauss Centre for Supercomputing e.V. (www.gauss-centre.eu) for funding this project by providing computing time on the GCS Supercomputer JUWELS and JUPITER at Jülich Supercomputing Centre (JSC). This research used resources of the Oak Ridge Leadership Computing Facility for the experiments on Quantinuum QPUs, which is a DOE Office of Science User Facility supported under Contract DE-AC05-00OR22725.

\bibliographystyle{IEEEtran}
\bibliography{references}

\end{document}